# DESIGN, ANALYSIS AND IMPLEMENTATION OF ELECTRONIC TEST FOR KNOWLEDGE EVALUATION IN THE COURSE OF INFORMATION TECHNOLOGIES FOR PHARMACEUTICAL STUDENTS

## Hristo Manev and Mancho Manev

Head Assistant Professor Hristo Manev, PhD

Medical University – Plovdiv, Faculty of Public Health, Department of Medical Informatics, Biostatistics and Electronic Education, hmanev@meduniversity-plovdiv.bg;

Professor Mancho Manev, PhD

Plovdiv University Paisii Hilendarski, Faculty of Mathematics and Informatics, Department of Algebra and Geometry, mmanev@uni-plovdiv.bg

AND

Medical University – Plovdiv, Faculty of Public Health, Department of Medical Informatics, Biostatistics and Electronic Education, mmanev@meduniversity-plovdiv.bg;

**Abstract:** The increased usage of the information technologies in everyday life and especially in education leads to demands for new forms of teaching, studying and appropriate examination and evaluation of acquired knowledge and skills of the students.

Modern electronic educational systems use only those technologies that improve the learning process and make it more effective. The interactive education provides an opportunity to develop skills for independent literature research and activation of the cognitive activity.

In this work, it is shown how a modern electronic education is implemented in the curriculum of English language pharmaceutical students at the Medical University – Plovdiv in the course of Information Technologies. It is developed a methodological approach of a hybrid system, i.e. a compulsory attendance at lectures in combination with two different types of conduction of the final test for comparison – a paper-based test and a remote web-based one. The results received from the parallel tests are processed and analysed and the conclusions are used to enhance the quality of the developed test and the type of implementation. Moreover, the examined students fill in an anonymous poll to show the authors their thoughts for this type of hybrid educational system.

## Introduction

The modern web-based e-learning systems relate only to those information technologies that improve the learning process. The effectiveness of education depends on many factors. One of the most important of them is the involvement of the students in the related course activities. This involvement can be achieved by students' interactive courses.

The interactive education provides many opportunities to develop the skills of the students. On the other hand, this considered type of teaching gives the opportunity for the implementation of interesting educational materials in order to increase the motivation and the interest of the students. Also, another positive result is the so-called two-way interaction teacher – student. This type of education can be for the both sides of the educational system a very valuable feedback and a teamwork booster.

The teaching and learning of the university discipline Information Technologies are a complex activity and many factors determine the success of it. The nature and quality of instructional materials, the presentation of the content, the pedagogic skills of the teachers, the learning environment and the motivation of the students are all important parts of the teaching-learning process and must be kept in view of any effort to ensure the quality of modern university education.

**Blended Teaching of IT at the Medical University – Plovdiv**

The so-called Educational Technology is a very wide field. Therefore we can find many definitions of it. The most popular are – Educational Technology is the usage of technology to improve education which is a systematic, iterative process for designing instruction which is used to improve the performance of the students; Educational Technology according to International Technology Education Association: Someone teaches with technology (uses technology as a tool); Concerned with the narrow spectrum of information and communication technologies; The fundamental goal of Educational Technology is to enhance the teaching and learning process.

Educational Technology has a dynamic character. The educational process starts with certain conditions. Permanent control of the lecturer is needed, stable reverse connection with the students and making competent decisions for managing the technological process. Furthermore, Educational Technology allows simultaneous usage of traditional and modern technologies. In the process of learning Information Technologies, the students are incessantly proposed their issues, research questions and solving problems in order to impress their teachers (Kenderov, 2010).

The so-called blended courses integrate face-to-face and online learning. Some of them use the online environment for their content or lecture delivery and the classroom for active learning opportunities. Others use the face-to-face time for lectures and the online environment for discussions, assessments and other learning. Another approach is to use a combination of these two, namely to employ the so-called "flipped classroom". In this type of teaching-learning process, the lectures are delivered online for some of the classes and students use them to prepare for active learning in the classroom. The in-class activities involve peer learning in small group activities to engage the students in problem-solving.

Nowadays the interest in online tests as a students' knowledge and skills evaluation has increased significantly. Some of the main reasons for this are: high efficiency of online tests, i.e. for a limited period of time the lecturers could assess a large number of students; opportunity for simultaneous testing of students from different specialties, courses and groups through various tests; the possibility of obtaining student's assessment immediately after completing the test; the teacher's ability to manage and expand the database of the tests' questions and the evaluation criteria.

**Model**

Looking to the aims of teaching Mathematics and Information Technologies, it can be seen that the important needed objectives underlying the mathematics subject are critical thinking, analytical thinking, logical reasoning, decision-making, problem-solving. Such objectives are difficult to be achieved only through verbal methods that are usually used. Therefore, in this work, it is proposed a model for the realisation of a hybrid educational site in Information Technologies for English language pharmaceutical students at the Medical University – Plovdiv, consisting of attendance hours for lectures and exercises, combined with the remote implementation of the test.

In the most of the universities in Bulgaria and abroad for these purposes it is mainly used the open-source system Moodle (Rice, 2006), (Cole, 2007), (Manev et al, 2012). There are also many Bulgarian university projects which are integrated into the studying process (Manev et al, 2014), (Rahnev et al, 2014).

**Implementation**

In this work, using the e-education system Moodle it is developed a methodological approach of a hybrid system, i.e. a compulsory attendance at lectures in combination with two different types of conduction of the final test for comparison – a paper-based test and a remote web-based one.

The main part of this project is the development and implementation of an electronic test for knowledge evaluation in the course of Information Technologies for pharmaceutical students – one of the most used forms of knowledge verification of the trainees. This form is used not only to test knowledge evaluation of the students but also for training by exercising their skills. The advantage of web-based teaching is that it can be done asynchronously, i.e. the educational information is available to students on-line and they can read it at a convenient time. The problem occurs when the student is required to perform tasks and tests. Online courses have difficulties in controlling fraud on tests. The lecturer could set a specific day and time for the test activity to contribute for the objectivity of the evaluation. Even if the student checked its answers in a textbook or online, the teacher can limit the time. So if there is not enough knowledge gained in their preparation the students are going to fail to end on time.

In the present work, it is designed an interactive test for the English language students studying the course "Information Technologies" at the Pharmaceutical Faculty of Medical University – Plovdiv.

The creation of a test in the e-system Moodle requires parameterization of list of different options related to the purpose of the examination (Figure 1 and Figure 2)

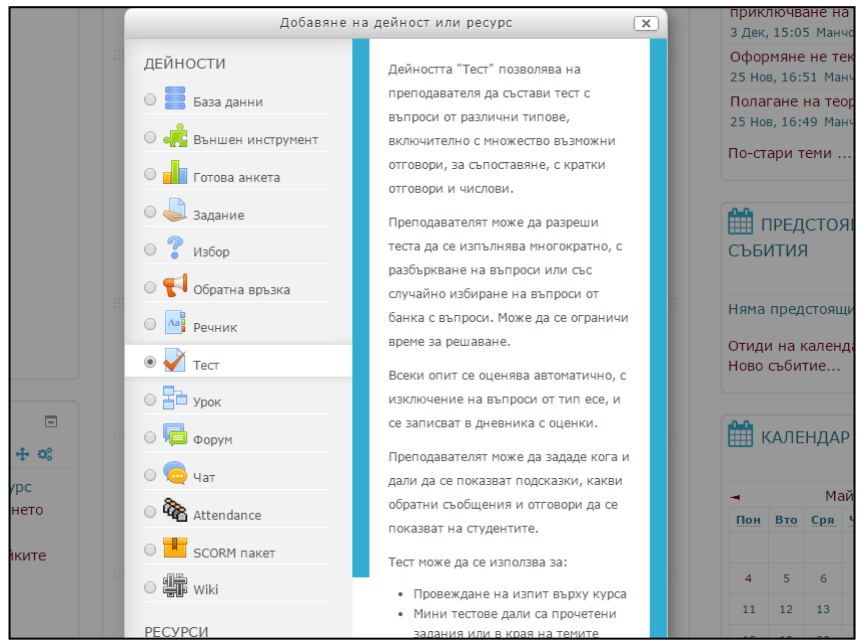

**Figure 1** Creating the test (Source: Authors)

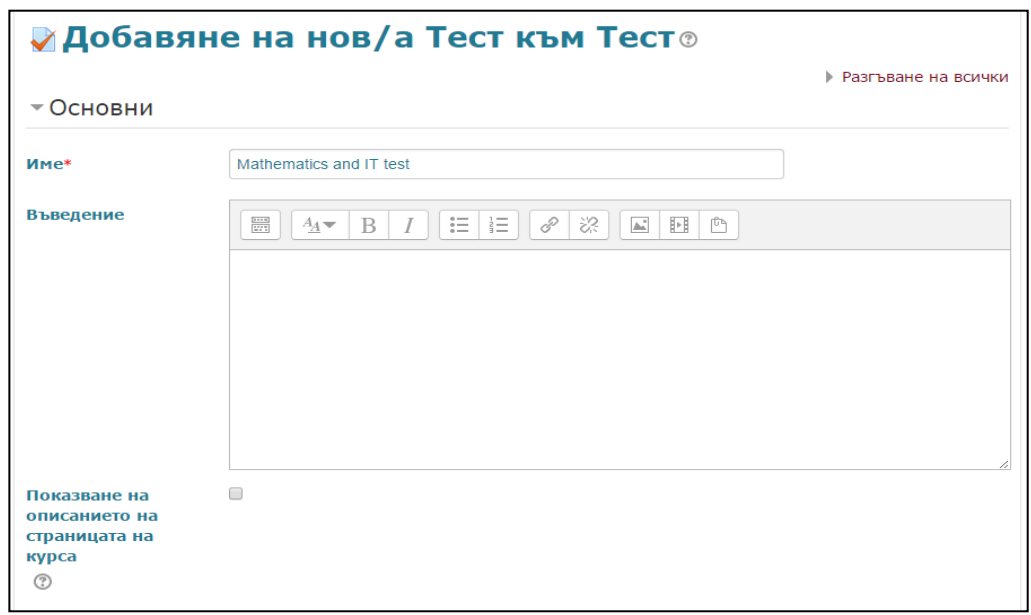

**Figure 2** Test parameterization (Source: Authors)

When the test is created and configured it is necessary to be created a database of questions that will be selected for the examination. Obviously, the larger database of questions makes more various tests and of course more accurate evaluation of the students (Figure 3).

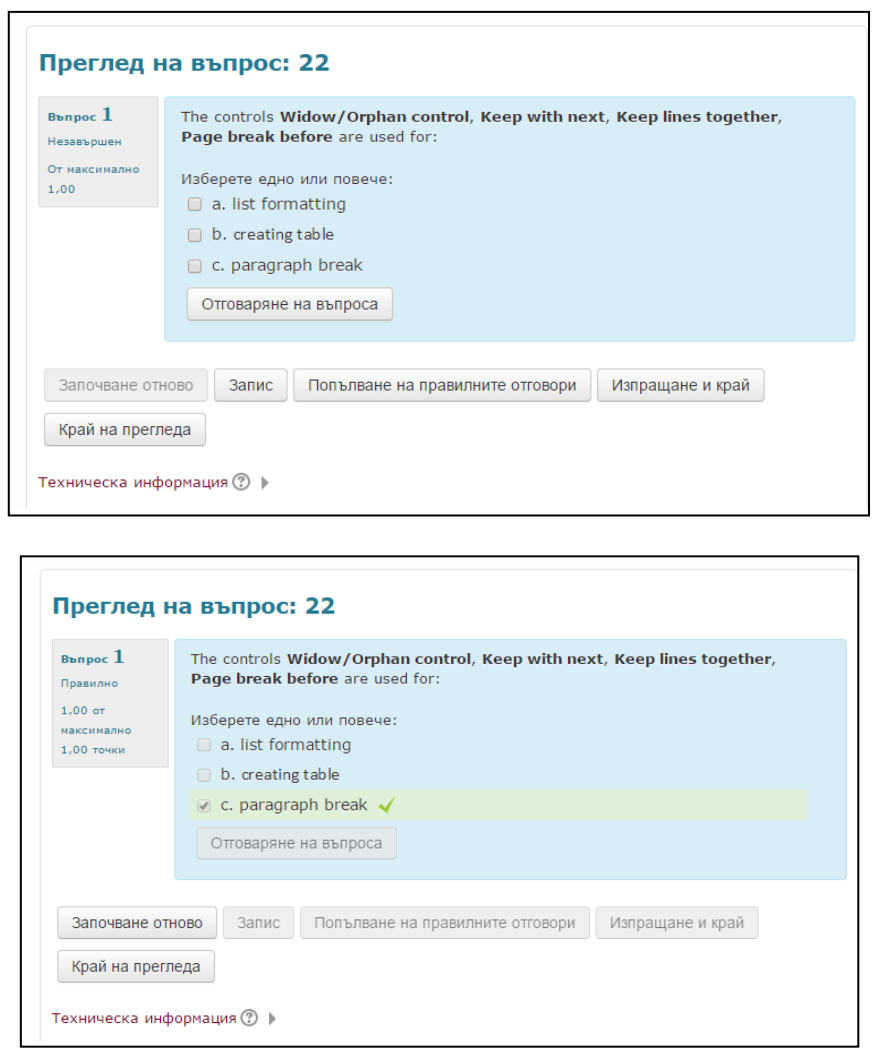

**Figure 3** Configuration of the questions (Source: Authors)

The tests are evaluated automatically and can be evaluated again if it is necessary. The questions can be open for the learners only for a certain period of time. After that, the students have not had access to the tests. After the completion of the tasks, the system allows comments or it can be shown the correct answers. The questions and the possible answers can be displayed in a different order to avoid cheating.

A great practical value has the submenu "Statistics". It provides a statistical analysis of the test. A dropdown menu allows the teacher to select which attempts of the students to be involved in the calculation of the statistics. The lecturer has the right to choose whether to display information about all attempts or just the first attempt. The full text of the statistic report can be downloaded in various formats. The Moodle system allows also more detailed analysis of each question individually (Figure 4).

## Analysis of responses

| Model response | Partial credit | Count | Frequency |
|---|---|---|---|
| или λ=0 или vec(_a_)=vec(_o_) | 100,00% | 22 | 56,41% |
| λ=0 | 0,00% | 6 | 15,38% |
| λ е перпендикулярно на vec(_a_) | 0,00% | 6 | 15,38% |
| (без отговор) | 33,33% | 4 | 10,26% |
| [No response] | 0,00% | 1 | 2,56% |

**Figure 4** Statistics (Source: Authors)

**Results**

For comparison, we implemented to the one and the same group of 10 students the electronic test and a paper-based one for knowledge assessment. The results received from the parallel tests are processed and analysed and the conclusions are used to enhance the quality of the developed test and the type of implementation. The evaluation is performed by a percentage scale (0-100%) corresponding to estimates by six-point scale (Weak (2) -Excellent (6)) (Figure 5).

| Student | Paper-based test | | Electronic test | |
|---|---|---|---|---|
| | Percentage | Assessment | Percentage | Assessment |
| Student №1 | 60,00 | Very Good (4,50) | 60,00 | Very Good (4,50) |
| Student №2 | 60,00 | Very Good (4,50) | 73,33 | Excellent (5,50) |
| Student №3 | 70,00 | Excellent (5,50) | 63,33 | Very Good (4,50) |
| Student №4 | 80,00 | Excellent (6,00) | 80,00 | Excellent (6,00) |
| Student №5 | 63,33 | Very Good (4,50) | 63,33 | Very Good (4,50) |
| Student №6 | 63,33 | Very Good (4,50) | 70,00 | Excellent (5,50) |
| Student №7 | 80,00 | Excellent (6,00) | 90,00 | Excellent (6,00) |
| Student №8 | 80,00 | Excellent (6,00) | 80,00 | Excellent (6,00) |
| Student №9 | 73,33 | Excellent (5,50) | 73,33 | Excellent (5,50) |
| Student №10 | 73,33 | Excellent (6,00) | 83,33 | Excellent (6,00) |
| **Average** | **70,33 – Excellent (5,50)** | | **73,67 – Excellent (5,50)** | |

**Figure 5** Two types of testing (Source: Authors)

**Conclusion**

The successful integration of e-learning and the shown good results and positive feedback from students are preconditions for the continuation of this form of teaching in the future. Moreover, the examined students fill in an anonymous poll to show the authors their thoughts for this type of hybrid educational system. Their responses will be taken into account in the future development of this type of education.

The relevance of e-learning environments today and the need of actualization of the educational courses are the main reasons for the usage and the integration of this form of teaching more and more. In this article, it is shown a model for an educational platform of a university discipline which can be applied also in other teaching courses. The results of the conducted experimental parallel two types testing show us that the e-learning is welcomed by the students nowadays.